\documentstyle[psbox]{WORKSHOP}  

\begin{document}

\title{Highly Ionized Warm Absorbers in AGNs: Simulations with the IXO Calorimeter}

\author{
F. Tombesi$^{1,2}$, M. Cappi$^2$, T. Yaqoob$^3$, J. Reeves$^4$ and G.G.C. Palumbo$^1$  
\\[12pt]  
%
$^1$  Dipartimento di Astronomia, Universit\`a di Bologna, Via Ranzani 1, I-04127 Bologna, Italy  \\
$^2$  INAF-IASF Bologna, Via Gobetti 101, I-40129 Bologna, Italy \\
$^3$  Department of Physics and Astronomy, Johns Hopkins University, Baltimore, MD 21218, USA \\
$^4$ Astrophysics Group, School of Physical \& Geographical Sciences, Keele University, Keele, Staffordshire ST5 5BG, UK\\
%
{\it E-mail(FT): tombesi@iasfbo.inaf.it} 
}

\abst{

We have performed several simulations in order to test the scientific capabilities of the IXO calorimeter, with particular emphasis on the detection of absorption lines in the 3--11~keV band. We derived the flux limits for their detection on several time-scales, compared different response matrices available and simulated realistic spectra from photo-ionized warm absorbers in AGNs. This study illustrates the considerable improvements that this instrument will bring to high resolution spectroscopy, especially related to the study of accretion and outflows in the central regions of AGNs. 

}

\kword{X-ray spectroscopy --- absorption lines --- AGNs}

\maketitle
\thispagestyle{empty}

\section{Introduction}

There is an increasing evidence for the presence of narrow blue-shifted absorption lines at rest-frame energies greater than 6~keV in the spectra of a number of radio-quiet AGNs (see review by Cappi 2006).
These features are commonly identified with Fe XXV and/or Fe XXVI K-shell absorption from a highly ionized (log$\xi \sim$ 2--4 erg~s$^{-1}$~cm) zone of circumnuclear gas, with column densities as large as $N_H\sim 10^{23-24}$~cm$^{-2}$.
The lines blue-shifts are also often quite large, reaching (mildly) relativistic velocities (up to 0.2--0.4c). In some cases short-term variability has been reported (e.g. Pounds et al.~2003; Reeves et al.~2004; Dadina et al.~2005; Markowitz et al.~2006; Braito et al.~2007; Cappi et al.~2009; Tombesi et al. in prep). 
These findings suggest the presence of previously unknown highly ionized and high velocity outflows from the central regions of radio-quiet AGNs, possibly connected with accretion disk winds/ejecta. 
The planned International X-ray Observatory (IXO) is a future large space observatory that will carry several instruments in the X-ray band. We concentrated on the X-ray Micro-calorimeter Spectrometer (XMS), which will provide a very high effective area ($\simeq$0.65~m$^2$ at 6~keV), coupled with a high energy resolution (FWHM$\simeq$2.5~eV) from $\sim$0.1~keV up to $\sim$12--13~keV.
Its unprecedented sensitivity will surely give a huge improvement to high resolution spectroscopy in the Fe K band.
Therefore, we performed XMS simulations in order to test the capabilities offered by this instrument regarding the detection of narrow absorption lines in the Fe K band. This is particularly important for the study of the highly ionized absorbers recently detected in the X-ray spectra of several radio-quiet AGNs.

\section{Narrow lines detection flux limits}

We derived the 2--10~keV flux limits for the 5$\sigma$ detection of narrow absorption lines in the 3--11~keV band of the XMS (the same results also apply to narrow emission lines). 
We assumed a typical AGN power-law continuum with $\Gamma=2$. The background has been modeled with two components: the internal non-X-ray background, parametrized with an energy-independent photon flux of $2\times 10^{-2}$~ph~cm$^{-2}$~keV$^{-1}$ and the contribution from unresolved AGNs (treated as in De Luca \& Molendi 2004). 
We used the core-glass response matrix\footnote{A list of different XMS response matrices is provided at http://ixo.gsfc.nasa.gov}, with a 2.5~eV (FWHM) resolution at all energies and 0.5~eV bin channels. 
The 2--10~keV flux limits for the detection of an absorption line with an equivalent width of 10~eV are reported in Fig.~1. We tested different exposure times, from 100~ks down to 10~ks and 5~ks, to check the lower limits for variability studies. Focusing in the region where Fe K absorption lines are more probably expected to be present ($\sim$6--9~keV), it will be possible to clearly detect lines with EW$=10$~eV (50~eV) in sources with 2--10~keV fluxes of $\sim 10^{-12}$ ($\sim 10^{-13}$) erg~s$^{-1}$~cm$^{-2}$ (for an exposure of 100~keV). Moreover, it will be possible to perform variability studies on time-scales as short as 5 (10)~ks for sources with 2--10~keV fluxes of $\sim 10^{-11}$ ($\sim 10^{-12}$) erg~s$^{-1}$~cm$^{-2}$.
We compared the narrow lines detection flux limits for two different proposed IXO calorimeter response matrices: the core-glass and the core-pore matrices, the latter considering also the loss in effective area due to the gratings. 
The main difference in the 3--11~keV band is a lower effective area for the core-pore, which results in an increase in the flux limits for the narrow line detection of about 60\% with respect to the core-glass.

\begin{figure}[t]
\centering
\psbox[xsize=6.5cm,ysize=6.5cm,rotate=r]{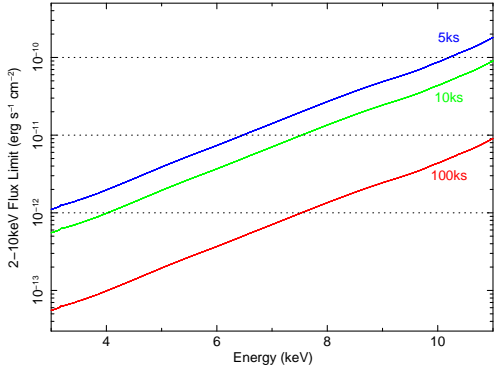}
\caption{ 2--10~keV flux limits for the 5$\sigma$ detection of narrow absorption lines with the XMS in the 3--11~keV band.}
\end{figure}

\section{Spectra simulations}

We carried out realistic spectral simulations of highly ionized and massive warm absorbers observed with the IXO calorimeter. We focused in the Fe K band, between 6~keV and 10~keV. 
The absorbers were modeled using the photo-ionization code Xstar and the parameter values were fixed to those typically observed in Seyfert galaxies (see Cappi 2006). 
We assumed a simple power-law SED with photon index of 2, ranging from the IR to the hard X-rays. 
Solar abundances were assumed. We used a high turbulent velocity value of 1000~km/s, as commonly assumed for such extreme absorbers. 
The 2--10~keV flux was considered to be that of relatively bright AGNs ($10^{-11}$~erg~s$^{-1}$~cm$^{-2}$). We used the core-glass response matrix and subtracted the proper background from a circular region of 5 arcsec radius. The exposure time was fixed to 100~ks.
The expected XMS spectrum from an absorber with total column density $N_H=10^{23}$~cm$^{-2}$ and ionization parameter log$\xi=3$~erg~s$^{-1}$~cm is showed in Fig.~2. For simplicity we considered a null outflow velocity. 
The vertical lines indicate the rest-frame energy location of the most intense expected Fe XXV (K$\alpha$ at 6.697~keV and K$\beta$ at 7.880~keV) and Fe XXVI (Ly$\alpha$ at 6.966~keV and Ly$\beta$ at 8.25~keV) resonant absorption lines. These narrow highly ionized Fe absorption lines are clearly visible in the spectrum with high significance. 
The high energy resolution of the XMS will allow to measure their centroid energy with unprecedented accuracy and will help to unambiguously set their identification and measure any velocity shifts. 
Moreover, the lines fine structure components will not be blended (for lower turbulent velocities) and also the most intense lines from a wide range of iron ions would be individually measured. 

\section{Conclusions}

The high effective area of the IXO calorimeter will allow not only to detect weak absorption (emission) lines in the 3--11~keV band but also to investigate their short time-scale variability, for sources with 2--10~keV fluxes as low as $\sim 10^{-12}$~erg~s$^{-1}$~cm$^{-2}$. Thanks to its high spectral resolution, it will be possible also to discriminate the lines from different Fe ionic species and measure with high accuracy their broadening and centroid energy. These characteristics are all of fundamental importance to bring real improvements in our understanding of high energetic phenomena, such as those related with the studies of accretion and outflows in the central regions of AGNs.

\begin{figure}[t]
\centering
\psbox[xsize=6.5cm,ysize=6.5cm,rotate=r]{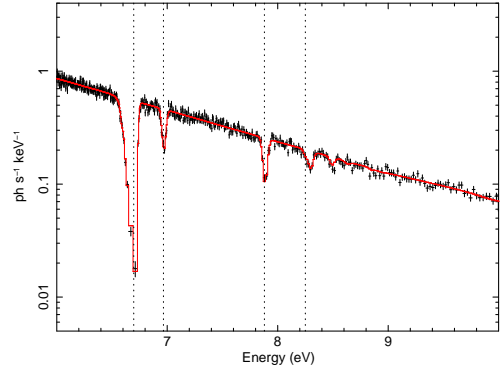}
\caption{Simulated spectrum of a highly ionized and massive absorber observed with the XMS in the 6--10~keV band.}
\end{figure}

\section*{References}

\re
Braito et al.~2007 ApJ, 670, 978
\re
Cappi 2006 AN, 327, 1012
\re
Cappi et al.~2009 2009arXiv0906.2438C
\re
De Luca \& Molendi 2004 A\&A, 419, 837 
\re
Markowitz et al.~2006 ApJ, 646, 783
\re
Pounds et al.~2003 MNRAS, 345, 705
\re
Reeves et al.~2004 ApJ, 602, 648

\label{last}

\end{document}